\documentclass[traditabstract]{aa} 
\usepackage{graphicx,epstopdf}
\usepackage{amsmath}
\usepackage{lscape}
\usepackage{longtable}
\usepackage{multirow}
\usepackage{multicol}
\usepackage{url}
\usepackage{natbib}
\usepackage{txfonts}
\usepackage{pdfpages}
\usepackage{longtable}

\def\nodata{ ~$\cdots$~}

\begin{document}
   \title{Outer-disk reddening and gas-phase metallicities: \\
   The CALIFA connection}
   
   \author{
   R.\,A.\,Marino\inst{1,}\inst{2}
	\and A.\,Gil de Paz\inst{3}
	\and S.\,F.\,S\'{a}nchez\inst{4}
	\and P.\,S\'{a}nchez-Bl\'{a}zquez\inst{5}
	\and N.\,Cardiel\inst{3,}\inst{6}
	\and A.\,Castillo-Morales\inst{3}
	\and S.\,Pascual\inst{3}
	\and J.\,V\'{i}lchez\inst{11}
	\and C.\,Kehrig\inst{11}
	\and M.\,Moll\'{a}\inst{7}
	\and J.\,Mendez-Abreu\inst{8}
	\and C.\,Catal\'{a}n-Torrecilla\inst{3}	
	\and E.\,Florido\inst{9,}\inst{10}
	\and I.\,Perez\inst{9,}\inst{10}
	\and T.\,Ruiz-Lara\inst{9,}\inst{10}
	\and S.\,Ellis\inst{12}	
	\and A.\,R.\,L\'{o}pez-S\'{a}nchez\inst{12,}\inst{13}
	\and R.\,M.\,Gonz\'{a}lez Delgado\inst{11} 
	\and A.\,de Lorenzo-C\'{a}ceres\inst{8}
	\and R.\,Garc\'{i}a-Benito\inst{11}
	\and L.\,Galbany\inst{14,}\inst{15}	
	\and S.\,Zibetti\inst{16}
	\and C.\,Cortijo\inst{11}
	\and V.\,Kalinova\inst{17}
	\and D.\,Mast\inst{18}
	\and J.\,Iglesias-P\'{a}ramo\inst{11,}\inst{19}
	\and P.\,Papaderos\inst{20,}\inst{21}
	\and C.\,J.\,Walcher\inst{22}
	\and J.\,Bland-Hawthorn\inst{23}
	\and the CALIFA Team\inst{19}\fnmsep\thanks{Based on observations collected at the German-Spanish Astronomical Center, Calar Alto, jointly operated by the Max-Planck-Institut f\"{u}r Astronomie Heidelberg and the Instituto de Astrof\'{i}sica de Andaluc\'{i}a (CSIC).}
          }

   \institute{
   CEI Campus Moncloa, UCM-UPM, Departamento de Astrof\'{i}sica y CC$.$ de la Atm\'{o}sfera, Facultad de CC$.$ F\'{i}sicas, Universidad Complutense de Madrid, Avda.\,Complutense s/n, 28040 Madrid, Spain, \email{ramarino@ucm.es}
	\and
	Department of Physics, Institute for Astronomy, ETH Z\"{u}rich, CH-8093 Z\"{u}rich, Switzerland 
	\and 
	Departamento de Astrof\'{i}sica y CC$.$ de la Atm\'{o}sfera, Facultad de CC$.$ F\'{i}sicas, Universidad Complutense de Madrid, Avda.\,Complutense s/n, 28040 Madrid, Spain
	\and
	Instituto de Astronom\'{i}a, Universidad Nacional Auton\'{o}ma de M\'{e}xico, A.P. 70-264, 04510, M\'{e}xico, D$.$ F$.$
	\and 
	Departamento de F\'{i}sica Te\'{o}rica, Universidad Aut\'{o}noma de Madrid, 28049 Madrid, Spain
	\and
	Instituto de F\'{i}sica de Cantabria (CSIC-Universidad de Cantabria), Avenida de los Castros s/n, E-39005 Santander, Spain
	\and
	CIEMAT, Departamento de Investigaci\'{o}n B\'{a}sica, Avda. Complutense 40, 28040 Madrid, Spain
	\and
	School of Physics and Astronomy, University of St Andrews, North Haugh, St Andrews, KY16 9SS, U.K. (SUPA)
	\and
	Instituto Universitario Carlos I de F\'{i}sica Te\'{o}rica y Computacional, Universidad de Granada, 18071 Granada, Spain
	\and
	Departamento  de  F\'{i}sica  Te\'{o}rica  y  del  Cosmos,  Facultad   de   Ciencias, Universidad de Granada, E-18071 Granada, Spain
	\and
	Instituto de Astrof\'{i}sica de Andaluc\'{i}a (IAA/CSIC), Glorieta de la Astronom\'{i}a s/n Aptdo. 3004, E-18080 Granada, Spain
	\and
         Australian Astronomical Observatory, PO Box 915, North Ryde, NSW 1670, Australia
        \and
         Department of Physics and Astronomy, Macquarie University, NSW 2109, Australia
         \and
	Millennium Institute of Astrophysics MAS, Nuncio Monse\~{n}or S\'{o}tero Sanz 100, Providencia, 7500011 Santiago, Chile
	\and
	Departamento de Astronom\'{i}a, Universidad de Chile, Camino El Observatorio 1515, Las Condes, Santiago, Chile
	\and
	INAF-Osservatorio Astrofisico di Arcetri - Largo Enrico Fermi, 5 -I-50125 Firenze, Italy
	\and
	Department of Physics 4-181 CCIS, University of Alberta, Edmonton AB T6G 2E1, Canada
	\and
	Instituto de Cosmologia, Relatividade e Astrof\'{i}sica, Centro Brasileiro de Pesquisas F\'{i}sicas, CEP 22290-180, Rio de Janeiro, Brazil
	\and
	Estaci\'{o}n Experimental de Zonas Aridas (CSIC), Carretera de Sacramento s/n, La Ca\~{n}ada, Almer\'{\i}a, Spain
	\and
	Instituto de Astrof\'{i}sica e Ci\^{e}ncias do Espa\c{c}o, Universidade do Porto, Portugal
	\and
	Centro de Astrof\'{i}sica da Universidade do Porto, Rua das Estrelas, 4150-762 Porto, Portugal
	\and
	Leibniz-Institut f\"{u}r Astrophysik Potsdam (AIP), An der Sternwarte 16, D-14482 Potsdam, Germany
	\and
	Sydney Institute for Astronomy, School of Physics A28, University of Sydney, NSW2006, Australia
	 }
 
   \date{Received July 2015; Accepted September 2015}
  
 \abstract
   {We study, for the first time in a statistically significant and well-defined sample, the relation between the outer-disk ionized-gas metallicity gradients and the presence of breaks in the surface brightness profiles of disk galaxies. SDSS {\it g'}- and {\it r'}-band surface brightness, {\it (g'}- {\it r')} color, and ionized-gas oxygen abundance profiles for 324 galaxies within the CALIFA survey are used for this purpose. We perform a detailed light-profile classification finding that 84\% of our disks show {\it down}- or {\it up-bending} profiles (\textsc{Type ii} and \textsc{Type iii}, respectively) while the remaining 16\% are well fitted by one single exponential (\textsc{Type i}). The analysis of the color gradients at both sides of this break shows a {\it U-shaped} profile for most \textsc{Type ii} galaxies with an average minimum {\it (g'}- {\it r')} color of $\sim$0.5\,mag and a ionized-gas metallicity flattening associated to it only in the case of low-mass galaxies. More massive systems show a rather uniform negative metallicity gradient. The correlation between metallicity flattening and stellar mass results in {\it p}-values as low as 0.01. Independently of the mechanism having shaped the outer light profiles of these galaxies, stellar migration or a previous episode of star formation in a shrinking star-forming disk, it is clear that the imprint in their ionized-gas metallicity was different for low- and high-mass \textsc{Type ii} galaxies. In the case of \textsc{Type iii} disks, a positive correlation between the change in color and abundance gradient is found (the null hypothesis is ruled out with a {\it p}-value of 0.02), with the outer disks of \textsc{Type iii} galaxies with masses $\leq$10$^{10}$\,M$_{\odot}$ showing a weak color reddening or even a bluing. This is interpreted as primarily due to a mass down-sizing effect on the population of \textsc{Type iii} galaxies having recently experienced an enhanced inside-out growth. Getting further insights into these correlations require of both larger samples and stellar metallicity measurements which will be possible with the new generation of IFS surveys.}
 \titlerunning {Breaks, colors and gas abundances within CALIFA}
 
\maketitle

\section{Introduction}
After the pioneering works on surface photometry of nearby galaxies by \citet{pat40}, \citet{devac59}, \citet{1968adga.book.....S} and \citet{free70}, it became accepted that galaxy disks follow an exponential light profile. The {\it inside-out} scenario of galaxy formation predict that the outskirts of a galaxy need longer times for their assembly resulting in an exponential decline of the radial light distribution and of the metal abundances \citep{1991ApJ...379...52W,1998MNRAS.295..319M}.
In the last two decades, especially with the advent of CCD imaging first and SDSS drift-scanning imaging more recently, we have learnt that the vast majority of nearby disks show breaks\footnote{Breaks and truncations are sometimes referred as different phenomena as explained in \citet{2012MNRAS.427.1102M}. In this study we will focus our attention on the innermost change in the SB profiles happening at $\mu_{r}$=22.5 mag/\arcsec$^{2}$}, which will be called {\it breaks} hereafter. Note that the radial position of this break should not be affected by inclination effect, as suggested by \citet{2009MNRAS.398..591S, 2014MNRAS.441.2809M} in their surface brightness (SB, hereafter) radial profiles after several scale lengths, and these can be either {\it down}- or {\it up-bending}. \citet{2005ApJ...626L..81E} and \citet{PT06} proposed a detailed classification of the different SB distributions in three general categories: (i) \textsc{Type i} (\textsc{Ti}) profiles that follow a single exponential law beyond the bulge area along all the optical extension of the galaxies, (ii) \textsc{Type ii} (\textsc{Tii}) profiles that present a double exponential law with a {\it down-bending} beyond the break radius, and (iii) \textsc{Type iii} (\textsc{Tiii}) profiles that exhibit an {\it up-bending} in the outer part. The observational results obtained at high redshift \citep{2004A&A...427L..17P, 2008ApJ...684.1026A} also suggest that breaks are present in distant galaxies and that, once formed, are long-lived \citep{2007MNRAS.374.1479G,2009ApJ...705L.133M}. This variety of radial morphologies have been tentatively explained by different mechanisms: Outer Lindblad Resonances \citep[OLR,][]{2012MNRAS.427.1102M, 2013ApJ...771...59M}, the presence of a bar \citep{2006ApJ...645..209D} or long-lived spiral arms \citep{2011MNRAS.412.1741S}, a shrinking of the star-forming disk \citep[Z15 hereafter]{2009MNRAS.398..591S,2015ApJ...800..120Z}, changes in the star-formation triggering mechanisms \citep{1994ApJ...435L.121E}, satellite accretion or the existence of a star formation (SF) threshold radius beyond which only stellar migration would populate the outer disk \citep[R08 hereafter]{2008ApJ...675L..65R}. The recent findings of a reddening in the optical broad-band colors for 39 \textsc{Tii} profiles \citep[B08 hereafter]{2008ApJ...683L.103B} have provided a fundamental piece of evidence to the actual scenario for the formation of galaxy disks and posed a challenge to these mechanisms. B08 also found a characteristic minimum color associated with these {\it U-shaped} color profiles. Such reddening in the optical colors is better explained as being due to a shrinking of the region of where SF has taken place over time (Z15) or to stellar migration (R08). In particular, a minimum in the luminosity-weighted age (and resulting optical colors) results naturally from the theoretical predictions of the stellar migration scenario \citep[R08; SB09]{2014A&A...570A...6S}. \\
A direct prediction of the somewhat naive {\it inside-out} disk-formation scenario, under the assumption of closed-box chemical enrichment, is the presence of a universal radial abundance/metallicity gradient in disk galaxies \citep{1989MNRAS.239..885M,2000MNRAS.312..398B,2011ApJ...731...10M,2014A&A...563A..49S}. This is indeed observed in most of late-type galaxies for both the gas and stellar populations \citep[e$.$g$.$,]{2014A&A...563A..49S, 2015arXiv150604157G} but is still under debate whether this abundance gradient is valid for all disk galaxies and at all radii \citep{2014A&A...570A...6S, 2015MNRAS.448.2030H}. 
On the other hand, not all theoretical models produce elemental abundance radial distributions as perfect exponential functions. In this regard, \cite{2015MNRAS.451.3693M} shows how the radial distributions of oxygen abundance for a sample of theoretical galaxies with different dynamical masses is better fitted by a curve instead of only one straight line. Their distribution results flatter in the inner disk, and flattens again in the outer regions of disks, mainly in the less massive galaxies. This behavior is a consequence of the ratio between the SF and the infall rates in each radial region, which, in turn, is defined by the surface stellar profile and the gas (molecular and diffuse) radial distributions. More interesting, although the surface brightness do not show any flattening associated to the oxygen abundances ones, colors also have an {\it U-shape} at the outer regions of disks especially for galaxies with masses similar to the Milky Way (MW).
In addition, several investigations both in our MW \citep{1996MNRAS.280..720V} and in nearby galaxies \citep{2009ApJ...695..580B,2011MNRAS.415.2439R,2011MNRAS.412.1246G,bres12,2012ApJ...754...61M, seba12} have reported a shallower oxygen abundance gradient (or flattening) in the outskirts, beyond $\sim$2 effective radii, R$_{\rm{eff}}$. In general, these deviations to the universal abundance gradient are explained in terms of variations of the in-situ gas density or effective SF history, the presence of a bar, or coincidence with the corotation radius. Recently, \citet{2013MNRAS.428..625S} showed a clear correlation between this minimum in the oxygen distribution and the corotation radii for 27 galaxies, but the mechanisms causing such different behaviors are not yet fully understood.\\
A fundamental question therefore arises from these results: are the breaks observed in the SB profiles and the flattening in the oxygen abundance gradients connected? In order to investigate the role of the ionized-gas metallicity on the nature of the observed changes in SB and colors we analyze these properties in a large sample of nearby disk galaxies from the CALIFA Integral Field Spectroscopy (IFS) survey. 

 \begin{figure*}[!t]
 \vspace{0.3cm}
 \hspace{1.1cm}
\resizebox{17.5cm}{!}{\includegraphics{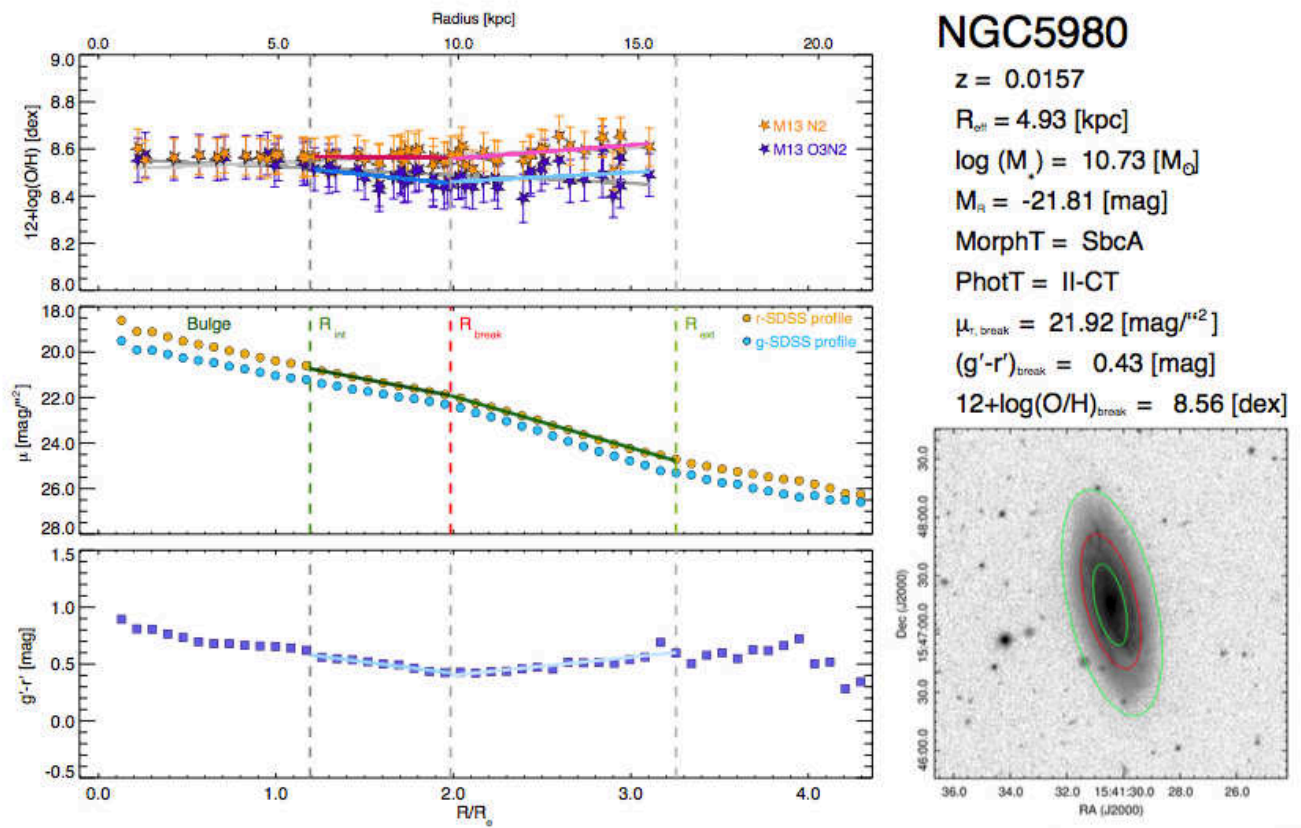}}
\vspace{-0.3cm}
   \caption{\footnotesize
   Radial ionized-gas metallicity, surface brightness and color profiles of the CALIFA disks: Example for galaxy NGC5980 showing a flattening beyond the break radius and the {\it U-shaped} color profile. {\it Left:} The radial oxygen abundance profile is shown in the top row: filled purple stars represent the metallicity values obtained for each \ion{H}{ii} region using the {\it M13-O3N2} calibration \citep{my13} while the filled orange stars show the ones obtained via the {\it M13-N2} calibration used in this work. The single fits of both profiles are drawn with grey solid lines. The double fits are plotted in red (inner part) and pink (outer part) for the {\it M13$-$N2} case and blue (inner part) and cyan (outer part) for the {\it M13-O3N2} one. The error bars plotted include both random and typical systematic errors associated to these calibrations. The SDSS SB profiles are shown in the second row: the {\it r'}-band profile is plotted with yellow circles and the {\it g'}-band one with cyan circles. The double fit performed to the {\it r'}-band is indicated with a green solid line. The third row shows the ({\it g'}$-${\it r'}) color gradient with violet squares and the double-linear fit is plotted with blue solid lines. The top x-axis shows the galactocentric radius in kiloparsecs while the bottom shows the galactocentric radius normalized at R$_{\rm{eff}}$. The vertical dashed (in color only in the second panel) lines mark the inner radius (green), the break radius (red) and the outer radius (green) used in each SB double fit. {\it Right:} The physical properties derived for each galaxy, including the SB type, are listed at right-top while the SDSS 3'$\times$\,3' {\it r'}-band post-stamp image is shown at right-bottom. The overplotted ellipses correspond to the same radii used for the SB double fit. (The complete atlas of the 324 CALIFA galaxies analyzed is available in the online version of this paper and also for convenience at \protect\url{https://guaix.fis.ucm.es/~raffaella/Thesis_ATLAS/CALIFA_ATLAS.pdf})}
\end{figure*}

\section{Data and Analysis}
\subsection{The Sample}
For this work we have selected the 350 galaxies observed by the CALIFA survey \citep{seba12a} at the CAHA 3.5m telescope with PMAS (Potsdam Multi Aperture Spectrograph) in the PPak mode \citep{2006PASP..118..129K} and processed by the CALIFA v1$.$5 pipeline \citep{rubenDR2} up to September 2014. CALIFA is an IFS survey, whose main aim is to acquire spatially resolved spectroscopic information of $\sim$600 galaxies in the Local Universe (0.005 $<{\it z}<$ 0.03), sampling their optical extension up to $\sim$2.5 R$_{\rm{eff}}$ along the major axis with a spatial resolution of FWHM~2.5" (1"/spaxel), and covering the wavelength range 3700\,\AA-7500\,\AA. By construction, our sample includes galaxies of any morphological type, being representative of all local galaxies between $-$23$<$\,M$_{abs,z}$\,$<$$-$18. Details on the data reduction are given in \cite{bernd13} and in \cite{rubenDR2}, and more information on the mother sample can be found in \citet{2014A&A...569A...1W}. We exclude from our analysis all those galaxies classified as mergers (26/350 galaxies), as interactions are expected to flatten the metallicity profiles independently of the secular mechanisms put to test in this work \citep[e$.$g$.$][S14]{2015arXiv150603819B, 2010ApJ...710L.156R}. Our surface-photometry sample therefore consists of 324 CALIFA galaxies. Global properties for the galaxies in our sample, such as morphological type, stellar mass, distance, etc., were taken from \citet{2014A&A...569A...1W}.

\begin{scriptsize}
\begin{table*}
\caption{Statistics of our sample according to the SB profiles classification. Quantities measured at R$_{\rm{break}}$ correspond to those galaxies where metallicity gradients could be measured (final sample). Errors represent the standard deviations.}\label{results}
\begin{center}
\begin{tabular}{l l | r | r r r | r r}
\hline\hline
\multicolumn{8}{c}{\textsc{Surface-photometry sample}} \\
\hline
\multicolumn{1}{l}{Quantities} &  \multicolumn{1}{l}{Units} & \multicolumn{1}{|c}{\textsc{Type i}} &  \multicolumn{3}{|c|}{\textsc{Type ii}} & \multicolumn{2}{|c}{ \textsc{Type iii}}\\
\hline
Number & {\small TOT=324} &  \multicolumn{1}{|c}{53} &  \multicolumn{3}{|c|}{172} &  \multicolumn{2}{|c}{99}\\
Frequency & {\small [\%] } & \multicolumn{1}{|c}{16.4} & \multicolumn{3}{|c|}{52.8} & \multicolumn{2}{|c}{30.8}\\
\hline
 & & &  \multicolumn{1}{c}{\textsc{Tii-CT}$^{\dagger}$}&  \multicolumn{1}{c}{\textsc{Tii.}o\textsc{-CT}$^{\dagger}$ }&  \multicolumn{1}{c|}{\textsc{Tii.}o\textsc{-OLR}$^{\dagger}$}&  \multicolumn{1}{c}{\textsc{Tiii-}{\small d}$^{\dagger}$ }& \multicolumn{1}{c}{ \textsc{Tiii-}{\small s}$^{\dagger}$}\\
\hline
Number & {\small [\#]} &  \multicolumn{1}{c|}{53} & 81 & 31 & 60 & 83 & 16\\
Frequency  & {\small [\%] } & \multicolumn{1}{c|}{16.4} & 25.0 & 9.6 & 18.5 & 25.6 & 4.9\\
$\mu_{0}$  & {\small [mag/$\arcsec$$^{2}$]} & 19.88$\pm$0.81 & 20.14$\pm$0.61 & 20.21$\pm$0.82 & 20.38$\pm$0.75 & 19.30$\pm$0.88 & 19.49$\pm$0.61\\
\hline
\multicolumn{8}{c}{\textsc{Final sample}} \\
\hline
Number & {\small TOT=131} & \multicolumn{1}{c|}{\nodata} & 37 & 18 & 43 & 30 & 3\\
Frequency  & {\small [\%] } & \multicolumn{1}{c|}{\nodata} & 28.2 & 13.7 & 32.8 & 23.0 & 2.3\\
R$_{\rm{break}}$ &  {\small [R$_{\rm{eff}}$] }& \multicolumn{1}{c|}{\nodata} & 1.43$\pm$0.48 & 1.43$\pm$0.37 & 1.47$\pm$0.38 & 1.50$\pm$0.49 & 1.50$\pm$0.47\\
$\mu_{\rm{break}}$ &  {\small [mag/$\arcsec$$^{2}$] } & \multicolumn{1}{c|}{\nodata} & 22.18$\pm$0.81 & 22.21$\pm$0.97 & 22.46$\pm$0.71 & 22.70$\pm$0.51 & 22.01$\pm$0.49\\
({\it g'}- {\it r'})$_{\rm{break}}$ & {\small [mag] } & \multicolumn{1}{c|}{\nodata} & 0.52$\pm$0.11 & 0.51$\pm$0.15 & 0.51$\pm$0.13 &  0.52$\pm$0.14 &  0.75$\pm$0.19\\
{\small (12+log(O/H))}$_{\rm{break,\,N2}}$ & {\small [dex/kpc] }& \multicolumn{1}{c|}{\nodata} &  8.50$\pm$0.08 & 8.46$\pm$0.11 & 8.52$\pm$0.08 & 8.51$\pm$0.09 & 8.58$\pm$0.09\\
\hline\hline
\end{tabular}
\end{center}
\hspace{0.5cm} $^{\dagger}$  {\small For a detailed explanation of each category see the classification schema presented in Fig$.$4 of \citet{PT06}. }\\ 
\end{table*}
\end{scriptsize}

\subsection{Surface brightness profiles and color gradients}
\label{SBandColor}
The SDSS {\it g'} and {\it r'} SB and ({\it g'}$-${\it r'}) color profiles were derived using the DR10 data products, in particular we use the {\it swarp} mosaicking code \citep{2014ApJS..211...17A}. We selected {\it g'} and {\it r'}$-$band data for two reasons: (i) they are deep enough to be sensitive to the outer part of galaxies and (ii) the breaks and corresponding {\it U-shaped} profiles were originally found in these SDSS bands (B08, Z15). We create 3'$\times$\,3' post-stamp images (as shown in Fig$.$1 for the galaxy NGC5980) and we estimate that our SB measurements are reliable up to 27$-$28\,mag/$\arcsec$$^{2}$ (note that DR10 images are sky subtracted contrary to the DR7 data used by B08 and that our faintest SB value for this analysis\footnote{This SB lower limit value ensures that our measurements are not affected by the contamination of the stellar halo which starts to contribute at fainter SB and at radii larger than 20 kpc \citep{2002AJ....123.1364W, 2010A&A...513A..78J, 2012MNRAS.419.1489B, 2013MSAIS..25...21B} neither the color are affected by the extended wings of the SDSS PSF.} is 27\,mag/$\arcsec$$^{2}$). For all galaxies in the sample we mask all contaminating sources such as bright stars or background galaxies and then we extract the flux in each band from the isophotal fitting provided by the IRAF task {\it ellipse}. Each isophote was computed varying both the ellipticity ($\epsilon$) and the position angle (PA) with a step of 1$\arcsec$. This approach should affect less the color profiles with respect to a procedure where $\epsilon$ and PA are kept fixed. The extracted fluxes were converted to AB magnitudes and corrected for Galactic extinction using the \citet{1998ApJ...500..525S} maps. Both {\it g'} (cyan circles) and {\it r'} (yellow circles) SB profiles are plotted in Fig$.$1 along with the resulting radial color profile. The details of the SB profile fitting procedure are given in Section~\ref{radprofs}.

\subsection{Oxygen abundance gradients}    
\label{metprofs}
We obtain spectroscopic information for $\sim$15130 \ion{H}{ii} regions (or complexes) from our 324 CALIFA datacubes using \textsc{HIIexplorer}\footnote{\textsc{HIIexplorer}: \url{http://www.caha.es/sanchez/HII_explorer/}}. Following the prescriptions described in S14 and the analysis presented in \citet{my13}, we compute the radial oxygen gradients for both N2 (log([N\,{\textsc{ii}}]$\lambda$6583/H$\alpha$)) and O3N2 (log(([O\,{\textsc{iii}}]$\lambda$5007/H$\beta$)/N2) indicators normalized at R$_{\rm{eff}}$ (see Fig$.$1). We refer to these calibrations as {\it M13-N2}\footnote{12 + log(O/H) = 8.743[$\pm$0.027]$+$0.462[$\pm$0.024] $\times$ N2} and {\it M13-O3N2}\footnote{12 + log(O/H)= 8.533[$\pm$0.012]$-$0.214[$\pm$0.012] $\times$ O3N2} hereafter, respectively. The disk effective radii (R$_{\rm{eff}}$) values for the galaxies analyzed in this work were taken from S14. For the current analysis, we use the results based on {\it M13$-$N2} as it provides a better match to the abundances obtained via {\it T$_{e}$}-based methods \citep{bres12,my13}. Instead of a single fit \citep[S14]{2014AJ....148..134P}, we perform two independent linear regressions in the inner and outer disk ranges to each side of the best-fitting SB break in the SDSS {\it r'}-band (see Section~\ref{radprofs}). This allows us to investigate whether or not there is a connection between SB and (O/H) breaks using a method that is less prone to the effects of outliers and the irregular sampling of the metallicity radial distribution provided by individual \ion{H}{ii} regions (compared to a direct double fit of the metallicities without priors). The oxygen abundance fits are computed including both systematic and random errors through Monte Carlo (MC) simulations. For each galaxy within our sample, we have performed 10$^{5}$ MC simulations to compute the difference in slopes and its uncertainty. We assume that the line fluxes are normally distributed according their estimated uncertainty, the metallicity has an intrinsic normal scatter of $\sigma$=0.0567 [dex] and the break radius is also normally distributed. This likely overestimates the uncertainties because part of the systematics in the {\it M13-N2} calibration might come from parameters that take the same value across the disk of a given galaxy but vary from galaxy to galaxy.

\begin{figure}[!t]
 \centering
\includegraphics[angle=90,width=0.8\hsize]{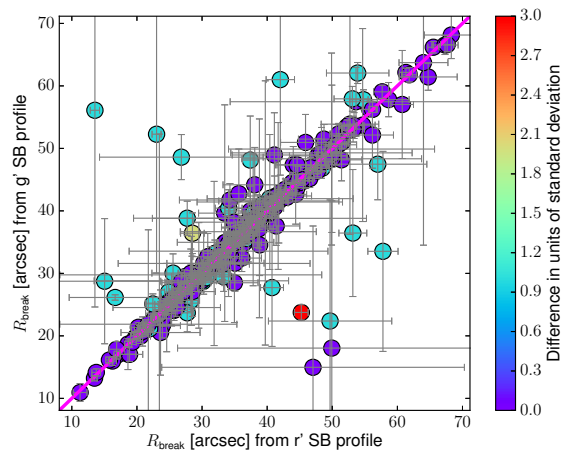}
   \caption{\footnotesize Comparison of the position of the break in the surface brightness profiles derived for the SDSS {\it g'} and {\it r'} bands for the total of 324 galaxies analyzed in surface photometry in this work. The color coding is shown at right and represents the offset from the 1:1 relation in units of the error of each individual point.}
\end{figure}
 
\section{Results}
\subsection{Radial profile classification}
\label{radprofs}

\begin{figure*}[!t]
 \centering
 \vspace{-0.7cm}
\resizebox{17cm}{!}{ \includegraphics[width=\hsize]{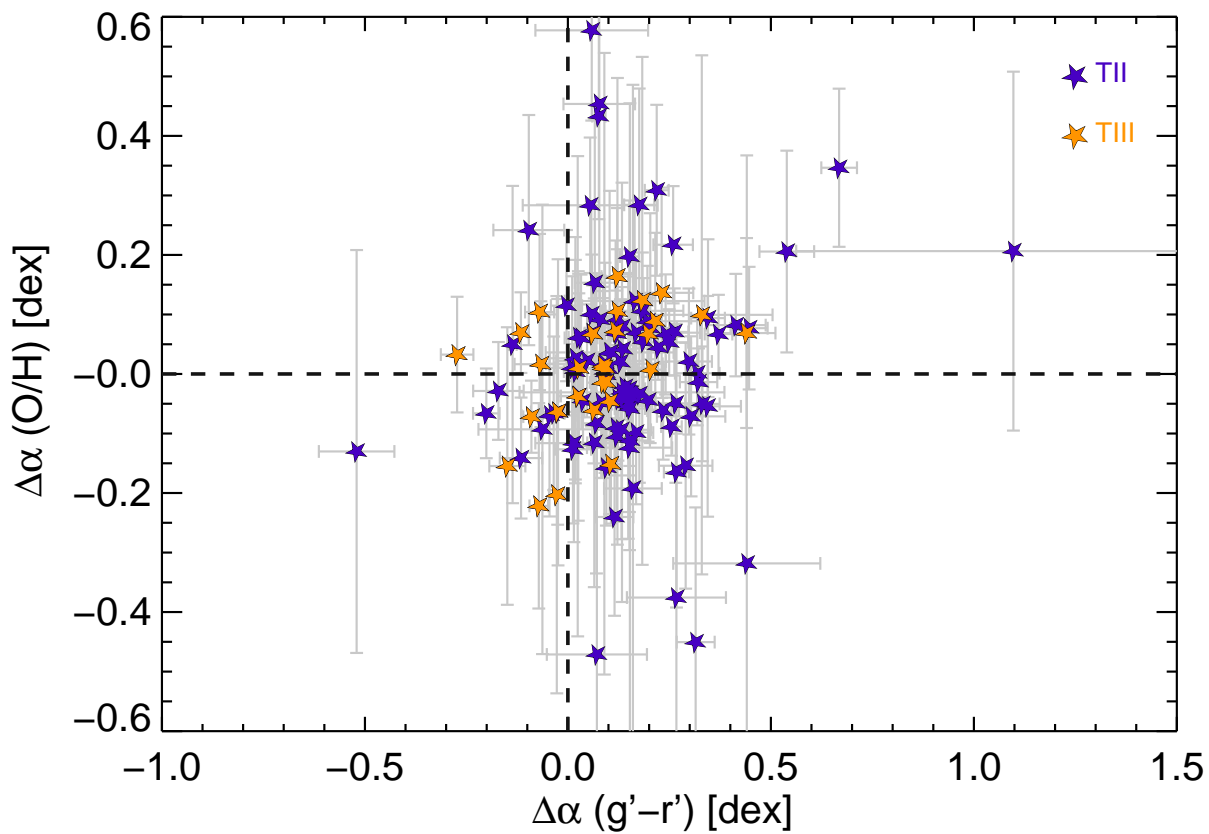}}
   \caption{\footnotesize Distribution of the oxygen metallicity slopes (outer-inner) difference, $\Delta$\,$\alpha$$_{\rm(O/H)}$, versus the {\it (g'}- {\it r')} color slopes one, $\Delta$\,$\alpha$$_{(g'- r')}$. We plot our \textsc{Tii} galaxies as purple stars while the \textsc{Tiii} are shown in orange.}
\end{figure*}

After excluding the bulge component, we have carried out a detailed analysis of the disk {\it g'}- and {\it r'}-band profiles. We identify the transition radius from the bulge to the disk in the profile (innermost point of our fitting range) as that where (1) there was an evident change in the isophote's ellipticity and (2) the brightness of the extrapolated inner disk component was equal or brighter than that of the bulge. Our procedure is aimed at deriving the broken exponential (including the position of the break radius) that best fits our SB profile via bootstrapping (see resulting best fits in the middle panel of Fig$.$1). Since the position of the break radii are found to be filter-independent (B08), our profile fitting and classification was performed on the {\it r'}-band data owing to their better S/N with respect to those on the {\it g'}-band (where the breaks appear brighter). In spite of that, as a consistency check we initially derived the position of the break in both bands independently finding a very good overall agreement between the two break radii (see Fig$.$2). We find that in the {\it r'}-band only 16\% of the CALIFA galaxies are well described by a single exponential law (\textsc{Ti} profiles) while the remaining 84\% of galaxies are better described by a broken exponential. In particular, 53\% of our disks present down-bending profiles and were classified as \textsc{Tii} and the remaining 31\% are \textsc{Tiii} (up-bending) profiles. Previous studies by \citet{2005ApJ...626L..81E} and \citet{PT06} (among others) have proposed that according to the presence of a bar or/and to the relative position of the break respect to the bar, the \textsc{Tii} class could be divided in different subgroups. The up-bending \textsc{Tiii} breaks represent 31\% of our sample and historically they are also subdivided according to the possible nature of the outer zone (spheroid or disk like outer region). Our statistical analysis is focused on the possible relation between the outer-disk reddening and the ionized-gas metallicity but is not aimed at explaining in detail the physical nature of each subgroup of the \textsc{Tii} and the \textsc{Tiii} categories. To achieve this task, in this study we will consider only the main classes for the join analysis of the stars and gas profiles to easy compare our results with previous findings.

Finally, the results of our disks classification are presented in Table~1, including the detailed frequencies and the SB, color and oxygen abundance measurements at break radius for each subtype. We conclude that our results are consistent with the previous classifications and that our breaks occur, as expected, at $\sim$2.5 scale-lengths (or $\sim$\,1.5\,$\times$\,R$_{\rm{eff}}$) on average and also that the mean ({\it g'}$-${\it r'}) color at R$_{\rm{break}}$ is similar to the one obtained by B08 for \textsc{Tii} disks ($\sim$0.5 mag, see Table~1),  although with a significant larger (> 2x) sample. Note that the latter color value is an average observed measurement so it has not be corrected for internal reddening, which could vary as a function of the galaxy inclination.

\subsection{The interplay between stellar light and abundance profiles}
The goal of this work goes beyond the disk classification of the CALIFA galaxies. Our main aim is to find possible connections between the stellar light colors and the gas metallicity in the external parts of disk galaxies. In order to ensure a good statistical sampling we impose that our final sample must include only spiral galaxies that have a minimum 5 \ion{H}{ii} regions beyond the break radius and also present a broken exponential light profile (i$.$e$.$ elliptical and \textsc{Ti} galaxies are excluded from the following analysis). The final sample comprises a total of 131 galaxies (98\,\textsc{Tii}\,+\,33\textsc{Tiii}) that fulfill these requirements and reduces the number of  \ion{H}{ii} regions used to 8653 from the 15130 detected in the surface photometry sample. We carry out two linear regressions in the same SB intervals to calculate the difference between the slopes of the outer-to-inner color profiles, $\Delta$\,$\alpha$$_{(g'- r')}$. As described in Section~\ref{metprofs}, we then apply the same analysis to the radial distribution of the oxygen abundance of \ion{H}{ii} regions and simultaneously fit the metallicity gradients within and beyond the {\it r'}-band R$_{\rm{break}}$ (obtaining the difference of the outer-to-inner oxygen slopes, $\Delta$\,$\alpha$$_{\rm(O/H)}$). We find a flattening or an inverted oxygen abundance trend beyond the break radius for 69/131 galaxies, which are the ones showing positive differences, $\Delta$\,$\alpha$$_{\rm(O/H)}$$\,>\,$0 (difference outer-inner). Negatives values of $\Delta$\,$\alpha$$_{\rm(O/H)}$ indicate a relative drop in the external part of the oxygen radial profile (as most profiles show a negative internal metallicity gradient).
The difference between the outer and the inner slopes ($\Delta$\,$\alpha$$_{\rm(O/H)}$) of our (O/H) fits is plotted in Fig$.$3 versus the color one, $\Delta$\,$\alpha$$_{(g'- r')}$. In this figure we represent the difference, $\Delta$\,$\alpha$$_{\rm(O/H)}$, of oxygen abundance slopes (outer$-$inner) versus the color slope difference, $\Delta$\,$\alpha$$_{(g'- r')}$, along with their errors obtained through the propagation of the fitting uncertainties. In general, our best-fitting results are in agreement with the oxygen abundance slope distributions obtained by \citet{2014A&A...563A..49S} and \citet{2015MNRAS.448.2030H}. In the case of the {\it M13-N2} our mean values for the inner and the outer slopes are $-$0.044 and $-$0.036 [dex/R$_{\rm{eff}}$], (median $-$0.041 and $-$0.029) respectively.

\begin{figure*}[!t]
 \centering
 \vspace{-0.5cm}
\resizebox{17cm}{!}{\hspace{-2cm}\includegraphics[width=0.98\hsize]{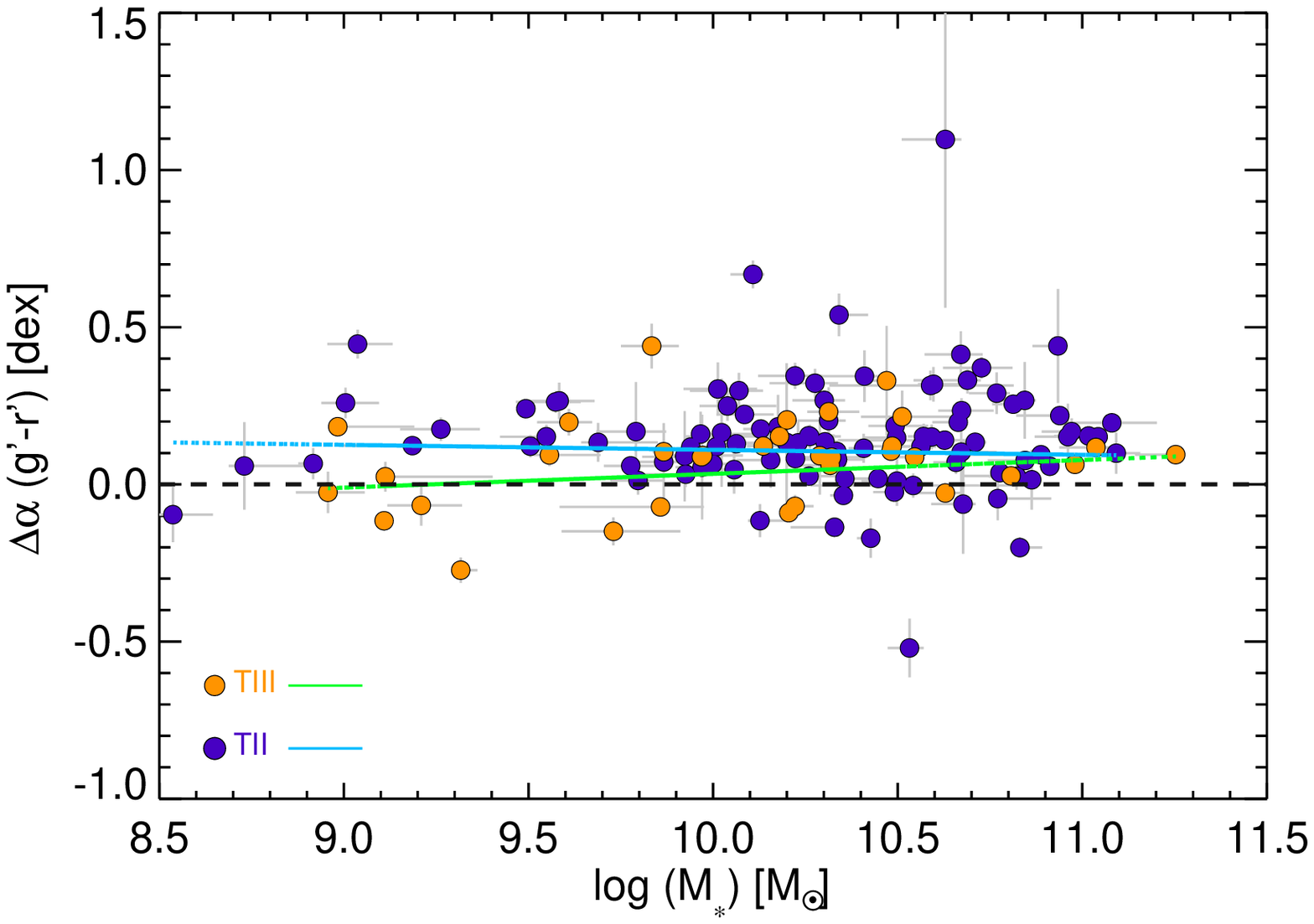}\hspace{2cm}\vspace{-0.1cm}
						       \includegraphics[width=0.98\hsize]{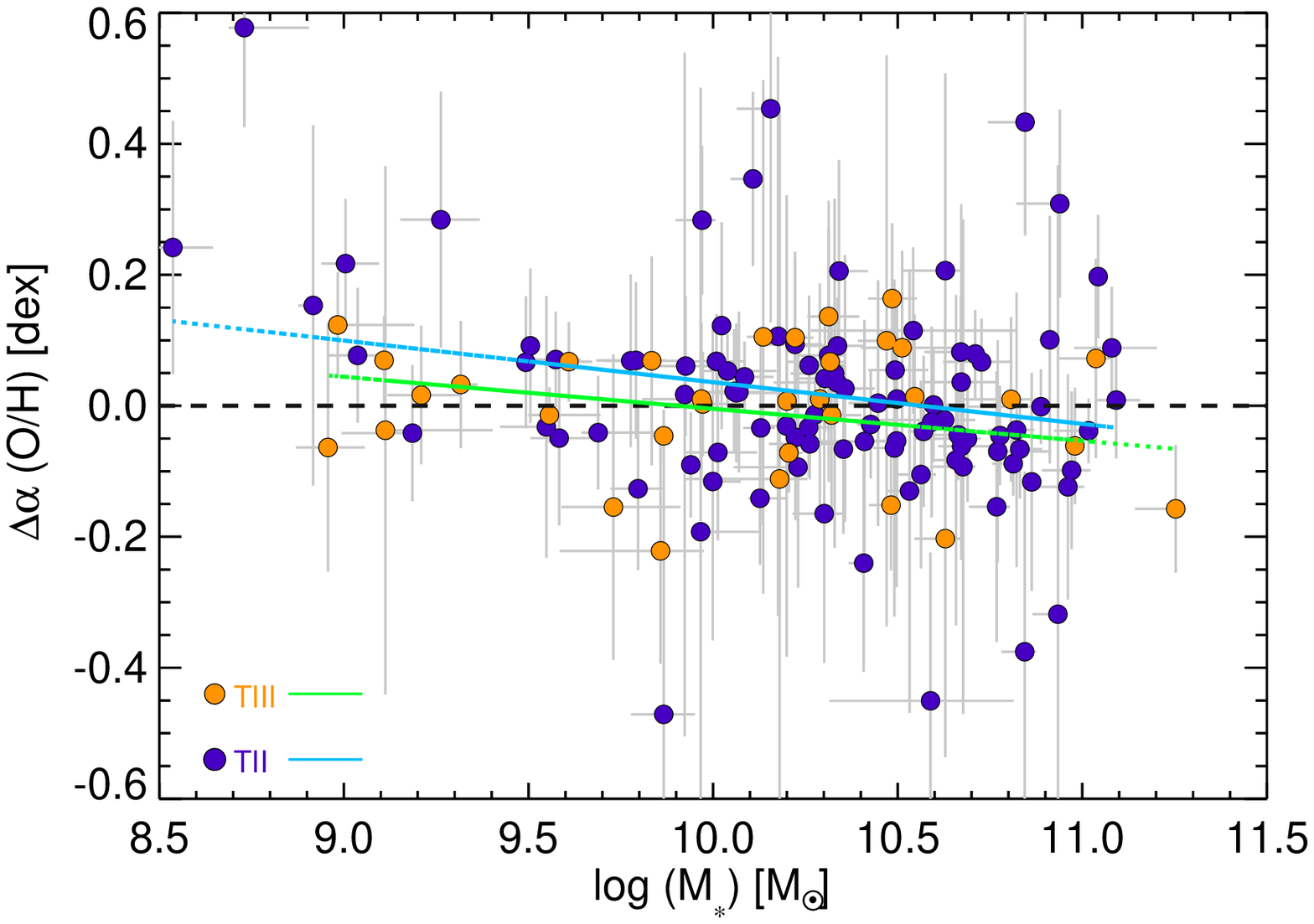}}
   \caption{\footnotesize
   {\it Left:} Change in the outer-disk color gradient as a function of the galaxy stellar mass (as derived by \cite{2014A&A...569A...1W}) along with the linear fits to the values derived for the \textsc{Tii} (cyan solid line) and \textsc{Tiii}-type (green solid line) galaxies. The {\it p}-value parameters in this case range between 0.1 and 0.2, although the dichotomy between \textsc{Tii} and \textsc{Tiii} at low stellar masses is quite clear. {\it Right:} The same as above but for the change in ionized-gas metallicity gradient. The correlation in the case of the \textsc{Tii} galaxies yields very low {\it p}-values, namely 0.003 (Pearson), 0.018 (Spearman), and 0.010 (Kendall) indicating that stellar mass might be one of the main drivers of the change in the ionized-gas metallicity in the outer disk of \textsc{Tii} galaxies.}
\end{figure*}

We find that 111/131 galaxies host color profiles that present a flattening or are {\it U-shaped}. In addition, 69 of these galaxies show also a change in the outer part of their oxygen gradients (50\,\textsc{Tii}\,+\,19\textsc{Tiii}). Our results suggest that {\it U-shaped} color profiles are more common in \textsc{Tii} than \textsc{Tiii} galaxies. However, although most \textsc{Tii} galaxies are barred or weakly barred galaxies, no correlation is found when we analyze the behavior of the whole sample with respect to the presence or not of a bar. When all galaxies are considered together, the probability of this distribution in difference of color gradient being the result of a galaxy population with no change in color gradient is smaller than 10$^{-4}$ ($i.$e$.$ none among the total of 10$^{4}$ MC runs used in this test). This result confirms the positive detection of the {\it U-shaped} color profiles within our sample. However, with regard to the change in metallicity gradient, while the distribution of gradient difference for \textsc{Tii} disks is compatible with the null hypothesis (the Kendall test $p$-value in this case is as high as 0.3), in the case of the \textsc{Tiii} disks there seems to be a correlation between color- and metallicity-gradient difference. Indeed, the $p$-value in this case for the null hypothesis is $<$0.02, so the presence of a positive correlation is supported by the data. The use of a statistical tests, including non-parametrical ones, are justified as the functional dependence between these properties is not known a priori. We have used three different tests for this purpose, namely the ones from Pearson, Spearman, and Kendall (see \cite{kendallgibbons} and references therein). All three tests yield very similar results in all correlations analyzed.  

\subsection{Mass dependence of metallicity and color breaks}
Despite \textsc{Tii} galaxies do not show a correlation between the change in color and metallicity gradient, it is worth analyzing whether these changes might be related with other properties of the disks either global or spatially resolved. The same can be said about the \textsc{Tiii}, where understanding the origin of the weak correlation between color and ionized-gas metallicity gradient would certainly require the analysis of other properties of these disks. Although splitting our samples of \textsc{Tii} and \textsc{Tiii} galaxies by physical properties would certainly benefit of an even larger sample of objects, we explore in this section the presence of potential correlations between the change in color and ionized-gas metallicity gradients with some global properties of our sample. 

Due in part to the reduced size of the sample once it is split in \textsc{Tii} and \textsc{Tiii}-type galaxies and also to the large uncertainty of the individual measurements of the change in color and metallicity gradient only the most significant of these potential correlations would stand out. In this regard, after analyzing the relation between these changes in color and metallicity gradients with (1) presence or lack of barred structures, (2) morphological type, or (3) galaxy stellar mass, only the latter is statistically significant within our sample, and only in the case of the \textsc{Tii} galaxies. In Fig$.$~4 we represent the change in color gradient (left panel) and the change in ionized-gas metallicity gradient (right panel) as a function of the stellar mass (as provided by \citet{2014A&A...569A...1W}). There is no clear dependence between the change in color gradient and the stellar mass, despite the obvious global reddening of the outer disks in our sample. The {\it p}-values derived are $\sim$0.9 (in all three tests carried out) in the case of the \textsc{Tii} galaxies and $\sim$0.4 in the case of the \textsc{Tiii} systems. However, in the case of the change in ionized-gas metallicity of \textsc{Tii} galaxies we find that more massive systems have a rather uniform negative metallicity gradient. than low-mass galaxies. Thus, at masses below 10$^{10}$\,M$_{\odot}$ a drop in the outer-disk metallicity gradient is commonly found. In this case the {\it p}-values found are as low as 0.003 (Pearson), 0.018 (Spearman), or 0.010 (Kendall). Again, a large number of systems, especially at the low-mass half of the distribution, would be desirable to confirm this relation. It is also worth noting that the outer-disk reddening is clear in \textsc{Tii} galaxies at all masses, while for \textsc{Tiii} galaxies, this is only clear at stellar masses above 10$^{10}$\,M$_{\odot}$. The segregation between \textsc{Tii} and \textsc{Tiii} galaxies below this stellar-mass value is very clear from this figure (left panel). 
The best linear fit (plotted in Fig$.$~4) in the case of the change in the color gradients as a function of stellar masses for \textsc{Tii} (\textsc{Tiii}) galaxies yields a slope of $-$0.016\,$\pm$\,0.006\, (0.044\,$\pm$\,0.005\,)[dex/M$_{\odot}$]. We also show in Fig$.$~4 the corresponding linear fit of the aforementioned correlation between the change in metallicity gradient and stellar mass in \textsc{Tii} (\textsc{Tiii}) galaxies in this case. The slope of the best fit is $-$0.06\,$\pm$\,0.02\,($-$0.05\,$\pm$\,0.02\,)[dex/M$_{\odot}$].

\section{Discussion}

\subsection{Outer-disk properties}
In order to interpret the nature of SB breaks in nearby galaxies we have carried out a joint analysis of CALIFA gas-phase metallicities and SDSS optical SB and colors. Whatever the mechanisms responsible for such breaks are, they should also be able to explain the diversity of morphologies, colors and metallicity gradients found in these, otherwise poorly understood, outskirts of disk galaxies. Moreover, any theoretical interpretation should also explain the results derived from this work (some of them already found by other authors), namely: \\
\noindent (\textsc{i}) The percentage of SB profiles and mean break colors found confirm those reported by previous works \citep{2005ApJ...626L..81E,PT06} B08, this time using the well-defined and large sample of nearby galaxies from the CALIFA IFS survey.\\
\noindent (\textsc{ii}) Most of the CALIFA \textsc{Tii} and \textsc{Tiii} disk galaxies show a flattening and even a reversed color gradients (see also B08).\\
\noindent (\textsc{iii}) The distribution of differences in the outer$-$inner (gas) metallicity gradient shows no correlation with the difference in color gradient in the case of the \textsc{Tii} disks, while there is a positive correlation between them (i$.$e$.$ a metallicity flattening) in the case of the \textsc{Tiii} disks.\\
\noindent (\textsc{iv}) The change in the ionized-gas metallicity gradient at both sides of the SB breaks in \textsc{Tii} disk galaxies varies with the galaxy stellar mass ({\it p}-value$\sim$0.01) in the sense that the low-mass galaxies show a more significant metallicity flattening (i$.$e$.$ with respect to the inner gradient) than more massive systems.\\
\noindent (\textsc{v}) At stellar masses below $\sim$10$^{10}$\,M$_{\odot}$, \textsc{Tii} and \textsc{Tiii} galaxies behave differently in terms of outer-disk reddening, with the latter showing little reddening or even a bluing in their color profiles. \\

Note that despite the evidence provided for the presence of these trends, the scatter is large. This suggests that each subgroup is rather inhomogeneous and therefore it likely includes galaxies with different spectro-photometric, chemical and dynamical histories.
A question naturally arises, can these observational results be reconciled in a unique disk formation theoretical scenario? 

\subsection{Outer-disk formation scenarios}
The level of detail reached by recent models of galaxy formation and evolution are finally allowing to use outer disks as laboratories for a better understanding of the relative contribution of in-situ SF, stellar migration and halo-gas and satellite accretion in shaping the observational properties of galaxy disks. Thus, \citet{2009MNRAS.398..591S} were able to estimate that 60\% of the stars in the outskirts of their simulated disk were not formed in-situ but migrated from the inner to the outer (warped) disk, leaving an important imprint on the stellar metallicity gradient. The idealized models of R08 indicate that the {\it U-shaped} and minimum in the color profile found by B08 is caused by a drop of gas surface density mainly due to changes in the angular momentum and that stars migrate mainly due to \textquotedblleft churning\textquotedblright\ effects \citep{2002MNRAS.336..785S}. The simulations of \citet{2012A&A...548A.126M} predict that secular processes (bars and spiral structures) could redistribute material towards several disk scale lengths (up to $\sim$10kpc). Z15 have recently proposed that most of the stars currently in the outer disks of a sample of galaxies observed with Pan-STARRS1 were not formed in-situ and the pollution of their outskirts is due to the combination of radial migration plus a truncation of the SF beyond the R$_{\rm{break}}$. These scenarios mainly differ in whether or not the effects of stellar migration dominate over those related with the time evolution of the size of the disk where star formation takes place. 

\subsection{Implications on the evolution of disks}
Our results indicate that the majority of our disk galaxies show {\it U-shaped} color profiles associated while more than half of them (69/131 disks in total present both features) have flat or inverted oxygen metallicity gradients. A correlation between the two is found but only in the case of the \textsc{Tiii} disks. \textsc{Tii} galaxies, on the other hand, where the outer-disk reddening is notorious, do not follow such a trend but when a metallicity flattening is present this becomes more severe as the stellar mass decreases. In this section we explore the implications (and constraints) of these results on the different theoretical scenarios proposed. 

Given the lack of correlation between the change in color and metallicity, and taken into account the typical sizes of our \textsc{Tii} disks, we infer that the change in metallicity associated to the observed color flattening cannot be larger than $\sim$0.4\,dex or that correlation should be present. Assuming such maximum change in metallicity we would expect a negligible change in the optical color of the stars associated to it. Thus, at a fixed age and SF timescale (from instantaneous to continuous) the change in ({\it g'}$-${\it r'}) between e.g$.$ 12+log(O/H)=8.3 and 8.7 would be smaller than $\sim$0.07\,mag (SB09, \citet{2014A&A...563A..49S}). So, our results indicate that metallicity alone (at least in the ionized-gas phase) would never explain the observed outer-disk color profiles of \textsc{Tii} galaxies. Despite the correlation found between the two quantities, even in the case of the \textsc{Tiii} galaxies, the amount of metallicity flattening does not seem to be enough to explain the reddening of their outer-disk optical colors. Besides, even if a stellar metallicity gradient would be present it is not obvious that could have an immediate effect on the ionized-gas phase abundances, especially in the case of the Oxygen, as this is released almost exclusively by short-lived massive stars. Therefore, we conclude that our results are in agreement with recent findings regarding positive age gradients in outer disks \citep{2009ApJ...697..361V, 2012ApJ...752...97Y}. In this regard, the work made by \citet{2015arXiv150604157G} on the stellar age radial profiles of 300 CALIFA galaxies (stacked by morphology and mass) has also shown a flattening of these profiles beyond 1.5-2  half light radius (HLR). \citet{2015arXiv150604157G} find negative extinction and stellar metallicity gradients, which leave the age as the only possible presumed responsible for the outer-disk reddening.

Therefore, any scenario aimed at explaining the color profiles presented in this work should also predict a radial change in the luminosity-weighted age of the stellar populations in outer disks, since the radial variation of either extinction \citep{2013A&A...556A..42H, 2015arXiv150604157G} or metallicity (see above) cannot. In principle, both the scenario where the radius of the disk where in-situ SF takes place shrinks with time (SB09) and the stellar migration scenario (R08) naturally predict a positive age gradient in the outer disks and are actually not mutually exclusive.  

We should note, however, that the use of ionized-gas metallicities could lead to more modest metallicity flattenings than those expected from the stars due to the dilution of the enriched gas by low-metallicity (or even pristine) gas from the halo \citep{2013ApJ...772..119L}, preferably in the outer parts of the disk. On the other hand, this could be compensated by the fact that this halo gas might have been previously polluted by metal-rich outflows originated during early phases of star formation in the disk \citep{bres12,2011MNRAS.416.1354D}. These inflows of unpolluted gas versus enriched inflows are always there, but certainly speculative if there is no clear evidence, e$.$g$.$ comparing the metallicity of the old population with the ionized gas metallicity. With the idea of overcoming those limitations, including also the potential contribution of satellite accretion to the population of the outer disks, a careful spectroscopic study of the stellar content in the outer parts of the CALIFA galaxies is being pursued by Ruiz-Lara et al. (in prep).

Our results indicate that the interpretation of the colors and ionized-gas metallicities of outer disks might be different for \textsc{Tii} and \textsc{Tiii} and, possibly, also for different stellar-mass ranges. The fact that virtually all \textsc{Tii} galaxies show a reddening in their outer-disk optical colors (independently of their stellar mass) already puts a clear difference compared with the \textsc{Tiii} galaxies (see below). Besides, we find that the metallicity flattening (although not correlated with the reddening in color) in \textsc{Tii} objects is more notorious at low stellar masses, something that is less clear in the case of \textsc{Tiii} galaxies. Finally, it is also worth keeping in mind that the mere shape of the \textsc{Tii} profiles indicate that the amount of stars found (whose presence is ought to be explained) beyond the SB break is smaller than that in the outer-disks of \textsc{Tiii} galaxies, at least for the intermediate-to-high stellar masses where \textsc{Tii} and \textsc{Tiii} galaxies show similar changes in their outer-disks color gradient (see left panel of Fig$.$4).

In the case of \textsc{Tii} galaxies we ought to explain (1) why, for a similar level of color reddening, the outer disks of low-mass systems show a more obvious metallicity flattening than high-mass ones but (2) being the age still the major driver of the radial change in color in either case. A possible explanation for the behavior observed in low-mass \textsc{Tii} galaxies is the presence of radial migration possibly due to the mechanism known as {\it churning} \citep{2002MNRAS.336..785S}, since these low-mass disks are expected to be kinematically cold (although some authors suggest that migration might be negligible in this case, \cite{2010ApJ...712..858G}). Unfortunately, current numerical simulations do not yet allow establishing whether this mechanism should lead to a larger radial metal diffusion but similar outer-disk color reddening than {\it heating}, that dominates the net stellar migration in more massive systems, once a large number of galaxies under different evolutionary conditions are considered (see e.g$.$ SB09). One aspect that should be taken into account when considering the feasibility of these migration mechanisms to explain the ionized-gas metal abundances of disks is the fact that the oxygen is virtually all released by massive stars, so the oxygen abundance of the ISM should be not altered by the presence of low-mass evolving stars that could migrate from the inner parts of the disks. However, since the oxygen abundances derived here rely on the intensity of the [\textsc{Nii}]6584\,\AA/H$\alpha$ line ratio and on the empirical relation between the N/O and O/H abundance ratios, a flattening in nitrogen abundance (which could be produced in this case by migrating intermediate-mass stars; see \cite{2013MNRAS.436..934W}) would also lead to an apparent flattening of the oxygen abundances derived. Here again, the comparison of ionized-gas and stellar abundances of outer disks could provide further clues. 

According to the scenario of a shrinking star-forming disk, the stellar population in the outer disks is mainly populated due to in-situ star formation. In that case we would expect that the drop in surface brightness would lead to a drop in the oxygen abundance (we are very close to the Instantaneous-Reclycing Approximation in this case) even if a positive color gradient is present in these regions, which is what we find for the most massive \textsc{Tii} galaxies. Should this scenario be valid for all \textsc{Tii} galaxies in general, we should be able to also explain why in low-mass \textsc{Tii} galaxies we find a signal of flattening in the oxygen abundance, despite the drop in surface brightness. Possible explanations could be that stellar migration is also playing a role in this case (see above) or that these galaxies have experienced episodes of extended star formation (which have led to the oxygen enrichment) on top of a secular shrinking of the size of the disk where star formation takes place during the long quiescent episodes (see also the case of \textsc{Tiii} disk galaxies below). \\

In the case of the \textsc{Tiii} galaxies, we find (1) a correlation between outer-disk reddening and ionized-gas metallicity flattening and (2) that galaxies with low level of reddening (or even bluing) are typically low-mass systems. These results are compatible with a scenario where low-mass \textsc{Tiii} galaxies are systems that have recently experienced (or are currently experiencing) an episode of enhanced inside-out growth, such as in the case of the Type-2 XUV disks \citep{2007ApJS..173..538T}, with blue colors and relatively flat metallicity gradients \citep{bres12}. In low-mass galaxies the small change in the metallicity gradient across the SB break would be consequence of their lower overall abundances and the presence of a rather homogeneous metallicity in outer disks. Indeed, recent cosmological hydrodynamical simulations by \citet{2011MNRAS.416.1354D} propose that accretion of IGM gas enriched by early outflows could be taking place in the outskirts of disks (see also \cite{2013ApJ...772..119L}). From the observational point of view, many are the results that show signs of accretion of metal-rich gas in the outer disks of spiral galaxies \citep{2015MNRAS.449..867B,2015MNRAS.450.3381L}. Finally, we cannot exclude that a fraction of the \textsc{Tiii} systems analyzed here could be also \textsc{Ti} disks (which are, indeed, also growing from inside out) with only a modest change in surface brightness at the break radius position.  

More massive \textsc{Tiii} galaxies, on the other hand, show a clear outer-disk reddening and corresponding metallicity flattening (through the correlation described above). This can be explained as due to the fact that they might have experienced episodes of enhanced inside-out growth (or, equivalently, XUV emission) in their outer disks in the past, that could have raised the oxygen abundance in these outer disks to the levels found in XUV disks \citep{bres12}, but where these have now decreased in frequency and/or strength. This is equivalent to a shrinking in the SF disk with time having occurred in the case of the massive ($\geq$10$^{10}$\,M$_{\odot}$) disks (see Z15 and references therein). In other words, our results indicate that the outer regions of spiral disks (at least the ones that are susceptible to have experienced outer-disk growth and, therefore, get classified as \textsc{Tiii}) also suffer from mass down-sizing effects. It would be worth exploring if this effect might be related to different gas fractions in the outer disks of these objects. Finally, whether stellar migration could be able to contribute significantly to the population of these shallow outer disks cannot be ruled out, at least in the case of the high-mass \textsc{Tiii} galaxies. The former interpretation, however, allows again to put all \textsc{Tiii} galaxies in the context of a common mass-driven evolutionary scenario. 

\section{Conclusions}
In this paper, we have explored the connections between the color and ionized-gas metallicity gradients in the external parts of the CALIFA disk galaxies. The main results of this paper are summarized as follows:
\begin{itemize}
\item We find a {\it U-shaped} color profiles for most \textsc{Tii} galaxies with an average minimum {\it (g'}- {\it r')} color of $\sim$0.5\,mag and a ionized-gas metallicity flattening associated in the case of the low-mass galaxies.
\item The distribution of differences in the outer$-$inner (gas) metallicity gradient shows no correlation with the difference in color gradient in the case of the \textsc{Tii} disks, while there is a positive correlation between them (i$.$e$.$ a metallicity flattening) in the case of the \textsc{Tiii} disks.
\item In the case of \textsc{Tiii} galaxies a positive correlation between the change in color and oxygen abundance gradient is found, with the low-mass \textsc{Tiii} ($\geq$10$^{10}$\,M$_{\odot}$) showing a weak color reddening or even a bluing.
\end{itemize}

Our view on the origin of these results in the context of the evolution of the outskirts of disks galaxies is:\\

\noindent (\textsc{i}) In the case of \textsc{Tii} galaxies, the observed color reddening could be explained by the presence of stellar radial migration.\\
\noindent (\textsc{ii}) Alternatively, within the scenario of a shrinking star-forming disk, these galaxies should have experienced episodes of extended star formation (which have led to the oxygen enrichment) on top of a secular shrinking of the size of the SF disk.\\
\noindent (\textsc{iii}) In the case of \textsc{Tiii} galaxies, a scenario where low-mass galaxies have recently shown an enhanced inside-out growth is proposed in order to explain the overall (negative) oxygen abundance gradient and the outer-disk bluing.\\
\noindent (\textsc{iv}) For more massive \textsc{Tiii} disks, the outer color reddening associated with a flattening in their oxygen gradients can be explained as due to a past inside-out growth that has now decreased in frequency and/or strength. Our results indicate that the outer regions of spiral disks also suffer from mass down-sizing effects.\\

Our results show that the CALIFA ionized-gas metallicities alone are not enough to tackle this aspects, furthermore deeper IFS data both for the stellar and the gas components as (MUSE, \citet{2010SPIE.7735E..08B}; MaNGA, \citet{2015ApJ...798....7B}; SAMI, \citet{2012MNRAS.421..872C}) should be analyzed in order to determine the relation between outer-disk (both gas and star) metallicity gradients and galaxy global properties, something that should allow establishing the mechanism(s) which dominate the photometric and chemical evolution of the outskirts of disk galaxies.
 
\begin{acknowledgements}
We are grateful to the anonymous referee for constructive comments and suggestions. R.A. Marino is funded by the Spanish program of International Campus of Excellence Moncloa (CEI). This study makes uses of the data provided by the Calar Alto Legacy Integral Field Area (CALIFA) survey (http://www.califa.caha.es). CALIFA is the first legacy survey being performed at Calar Alto. The CALIFA collaboration would like to thank the IAA-CSIC and MPIA-MPG as major partners of the observatory, and CAHA itself, for the unique access to telescope time and support in manpower and infrastructures. The CALIFA collaboration thanks also the CAHA staff for the dedication to this project. We thank Carmen Eliche-Moral and Antonio Cava for stimulating discussions at several points in the developments of this work. We acknowledge support from the Plan Nacional de Investigaci\'{o}n y Desarrollo funding programs, AyA2010-15081, AyA2012-30717 and AyA2013-46724P, of Spanish Ministerio de Econom\'{i}a y Competitividad (MINECO), as well as to the DAGAL network from the People's Program (Marie Curie Actions) of the European Union's Seventh Framework Program FP7/2007-2013/ under REA grant agreement number PITN-GA-2011-289313. C$.$C$.$$-$T$.$ thanks the support of the Spanish Ministerio de Educaci\'{o}n, Cultura y Deporte by means of the FPU fellowship program. CJW acknowledges support through the Marie Curie Career Integration Grant 303912. Support for LG is provided by the Ministry of Economy, Development, and Tourism's Millennium Science Initiative through grant IC120009, awarded to The Millennium Institute of Astrophysics, MAS. LG acknowledges support by CONICYT through FONDECYT grant 3140566. SFS thanks the CONACYT-125180 and DGAPA-IA100815 projects for providing him support in this study. JMA acknowledges support from the European Research Council Starting Grant (SEDmorph; P$.$I$.$ V$.$ Wild). PP is supported by FCT through the Investigador FCT Contract No. IF/01220/2013 and POPH/FSE (EC) by FEDER funding through the program COMPETE. He also acknowledges support by FCT under project FCOMP-01-0124-FEDER-029170 (Reference FCT PTDC/FIS-AST/3214/2012), funded by FCT-MEC (PIDDAC) and FEDER (COMPETE).

\end{acknowledgements}

\bibliographystyle{aa}
\bibliography{referencias}

\end{document}